\documentclass[aps,prl,preprint,superscriptaddress,amsmath,amssymb]{revtex4-1}

\usepackage{graphicx}
\usepackage{dcolumn}
\usepackage{bm}

\begin{document}

\title{Nano-scale displacement sensing based on Van der Waals interaction}

\author{Lin Hu}%
\affiliation{ICQD/Hefei National Laboratory for Physical Sciences at Microscale, and Key Laboratory of Strongly-Coupled Quantum Matter Physics, Chinese Academy of Sciences, and Department of Physics, University of Science and Technology of China, Hefei, Anhui 230026, China}

\affiliation{Synergetic Innovation Center of Quantum Information $\&$ Quantum Physics, University of Science and Technology of China, Hefei, Anhui 230026, China}

\author{Jin Zhao}%
\email{zhaojin@ustc.edu.cn}
\affiliation{ICQD/Hefei National Laboratory for Physical Sciences at Microscale, and Key Laboratory of Strongly-Coupled Quantum Matter Physics, Chinese Academy of Sciences, and Department of Physics, University of Science and Technology of China, Hefei, Anhui 230026, China}

\affiliation{Synergetic Innovation Center of Quantum Information $\&$ Quantum Physics, University of Science and Technology of China, Hefei, Anhui 230026, China}

\author{Jinlong Yang}
\email{jlyang@ustc.edu.cn}
\affiliation{ICQD/Hefei National Laboratory for Physical Sciences at Microscale, and Key Laboratory of Strongly-Coupled Quantum Matter Physics, Chinese Academy of Sciences, and Department of Physics, University of Science and Technology of China, Hefei, Anhui 230026, China}

\affiliation{Synergetic Innovation Center of Quantum Information $\&$ Quantum Physics, University of Science and Technology of China, Hefei, Anhui 230026, China}

\date{\today}

\begin{abstract}
We propose the nano-scale displacement sensor with high resolution for weak-force systems could be realized based on vertical stacked two-dimensional (2D) atomic corrugated layer materials bound through Van der Waals (VdW) interaction. Using first-principles calculations, we found the electronic structure of bi-layer blue phosphorus (BLBP) varies appreciably to both the lateral and vertical interlayer displacement. The variation of electronic structure due to the lateral displacement is attributed to the changing of the interlayer distance $d_{z}$ led by atomic layer corrugation, which is in a uniform picture with vertical displacement. Despite different stacking configurations, the change of in-direct band gap is proportional to $d_{z}^{-2}$. This stacking configuration independent $d_{z}^{-2}$ law is found also works for other graphene-like corrugated bi-layer materials, for example MoS$_2$. By measuring the tunable electronic structure using absorption spectroscopy, the nano-scale displacement could be detected. BLBP represents a large family of bi-layer 2D atomic corrugated materials for which the electronic structure is sensitive to the interlayer vertical and lateral displacement, thus could be used for nano-scale displacement sensor. Since this kind of sensor is established on atomic layers coupled through VdW interaction, it provides unique applications in measurements of nano-scale displacement induced by tiny external force.
\end{abstract}


\maketitle

High-resolution displacement sensing is essential for scientific measurement with a wide range of applications, among which, the nano-scale positioning measurement is becoming increasingly important in ultra-precision manufacturing and metrology in nano-technological applications. On micro-scale level, many micro-sensors such as microphones, accelerometers and pressure sensors, ultimately rely on accurately detecting small displacements with micro-scale resolution \cite{1,2,3}. With the development of nanotechnology, the transducer fundamental concepts at the micro-scale are extended into the nano-scale region \cite{4}. Nano-scale displacement sensing with high resolution has urgent applications for nano-devices.

Conventional displacement sensing are based on optical interferometry \cite{5}, capacitance \cite{6}, and piezoresistance \cite{7,8}. For the nano-scale sensing, the conventional methods encounters different drawbacks. The optical methods are limited by the drawbacks of non-linearity, fringe effects, proportionality errors, volume and cost of optical components, and mostly important, the light wavelength. Capacitive sensing suffers from signal loss and is limited by area \cite{6}. Piezoresistors scaled down to nano-scale leads to strongly increased resistance and the resistance noise followed \cite{7,8}. To achieve the high resolution in nano-scale, new materials, for example nanotubes, have been applied for nano-scale sensing \cite{4,9,10,11,12,13}. Yet despite these attempts in nano-materials, it is still complex to get high resolution for displacement induced by tiny force. For example, to measure the displacement of a quantum mechanical oscillator, a single electron transistor was applied as a displacement sensor to achieve the quantum-limited sensitivity \cite{1}. Hence the alternative complementary methods of measuring precise displacement for weak-force system have urgent need in both scientific research and practical applications.

Vertical stacked two-dimensional (2D) atomic layers formed by Van der Waals (VdW) interaction arouse intense research interest recently. The weak VdW interaction leads to many new physical phenomena including new van Hove singularities \cite{14,15,16,17}, Fermi velocity renormalization \cite{18,19} and Hofstadter＊s butterly pattern \cite{20,21,22,23,24} in graphene bi-layer as well as the tunable interlayer coupling in bi-layer semiconducting MoS$_2$ \cite{25}. In this Letter, we propose a new nano-scale displacement sensor based on bi-layer blue phosphorus (BLBP). Blue phosphorus is a new two dimensional (2D) material predicted by first principles calculations \cite{26}. Similar with other 2D layered materials, BLBP is coupled by VdW interaction. Using first-principles calculations within density functional theory (DFT), we found the electronic structure of BLBP is significantly sensitive to the lateral and vertical displacement between the two layers. In a uniform picture with vertical displacement, the sensitivity of electronic structure to the lateral displacement is due to the variation of the vertical interlayer distance ($d_z$) led by atomic layer corrugation. In spite of different stacking configurations, the change of in-direct band gap is proportional to $d_{z}^{-2}$. By detecting the change of electronic structure using absorption spectroscopy, one can measure the nano-scale displacement. This proposed sensor is based on VdW interaction, therefore, it provides unique applications in weak-force systems.

First-principles calculations are performed with the Vienna Ab-initio Simulation Package (VASP) \cite{27,28}. The Perdew-Burke-Ernzerhof (GGA-PBE) \cite{29} functional with VdW correction proposed by Grimme (DFT-D2) \cite{30} is used to optimize the structures and get the initial electronic structures \cite{31,32,33,34,35,36,37,38,39,40,41,42}. An energy cutoff of 600 eV is selected for the plane wave basis sets. The projector-augmented wave (PAW) \cite{43,44} method is used for electron-ion interaction. A vacuum spacing of about 15{\AA} ensures that the interactions between the layers are negligible. The dipole correction is applied to compensate the dipole interaction in the direction perpendicular to the plane. The k-point sampling uses the Monkhorst-Pack scheme \cite{45} and employs a $11\times11\times1$ mesh. The criterion of maximum force during optimization on each atom is less than 0.01 eV/{\AA} and the convergence criteria for energy is $10^{-5}$ eV. We scan the size of unit cell for each system to obtain the lattice parameters with the lowest energy. The spin-orbital coupling effect is included in the calculations. It is well known that DFT usually underestimated the band gap of semiconductor. To correct this problem, we chose HSE \cite{46} functional to obtain all the electronic structure information.

BLBP is a typical VdW bi-layer 2D system. The binding energy calculated by DFT-D2 is 25 meV/atom with a vertical interlayer distance as 3.22{\AA} for an A-A stacking bi-layer, which is at the same magnitude with graphene. The novelty for this bi-layer semiconductor is that the band structure significantly depends on the stacking configuration of the two BP layers. Fig.1 (b) and (c) give plottings of the in-direct band gap ($E_{gap}$) and total energy change ($E_{total}$) when we move the top layer BP laterally into different directions while keeping the bottom layer fixed as shown in Fig.1 (a). One can see that the band gap changes from 1.75 eV to 2.33 eV while the total energy changes within 12 meV. In Fig.1(c) the energies of valance band maximum (VBM) and conduction band minimum (CBM) respected to the vacuum level are plotted. It is shown that CBM is sensitive to the lateral displacement while VBM almost remains unchanged.

\begin{figure}
\includegraphics[width=\textwidth]{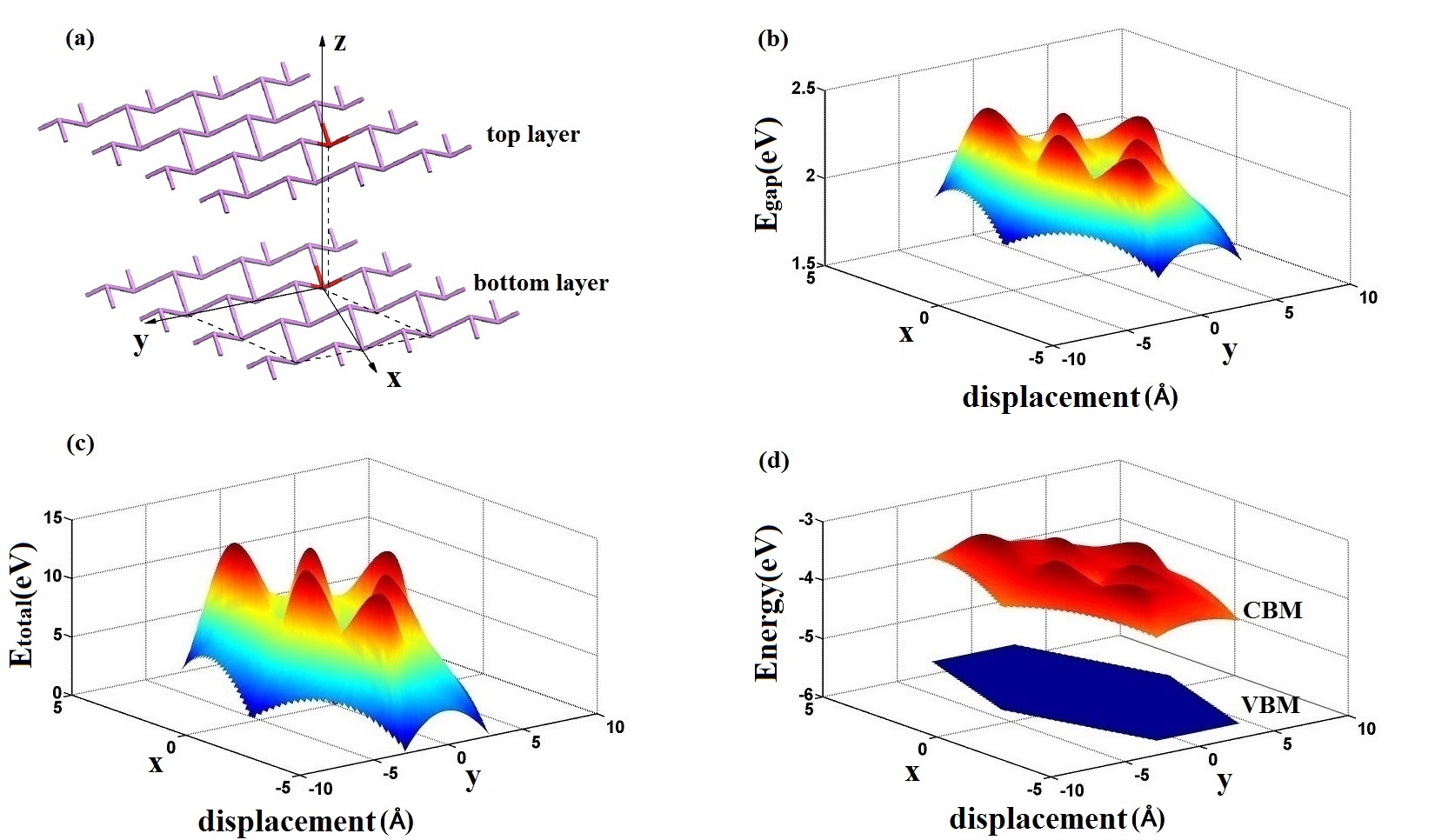}
\caption{(color online)The BLBP electronic structure dependence on the lateral displacement. (a) The schematic diagram of lateral interlayer displacement in which the bottom layer is fixed and top layer is moved laterally. (b-d) Three dimensional plotting of $E_{gap}$ (b), $E_{total}$ (c), $E_{VBM}$ and $E_{CBM}$ (d) when the top layer is moved laterally in $x$ and $y$ directions. For $E_{VBM}$ and $E_{CBM}$ the energy of vacuum level setting to zero.}
\end{figure}

We chose the $x$ direction in which $E_{gap}$ changes the most significantly to investigate the detailed electronic structure. Fig.2 (a) shows the three representative stacking configurations marked as A-A, A-B and A-B' when the upper layer moves along $x$ direction. $E_{gap}$ changes from 1.75 eV for A-A stacking to 2.04 eV for A-B stacking and then to 2.33 eV for A-B' stacking as shown in Fig.2(b). Consistent with Fig.1(d), the band structure plottings in Fig.2 (b) also show that the remarkable band gap change is mainly induced by the change of the conduction band (CB). The valance band (VB) dispersion also varies however the energy of VBM remains unchanged. To understand this, in Fig.2 (c) we plot out the spatial probability density distribution of VBM and CBM of BLBP with different stacking. It is clearly shown that for VBM, there is almost no interlayer interaction. Yet CBM is shown to be the 考 bonding state between two layers which contains significant delocalized orbital probability density in the interlayer region.

\begin{figure}
\includegraphics[width=0.5\textwidth]{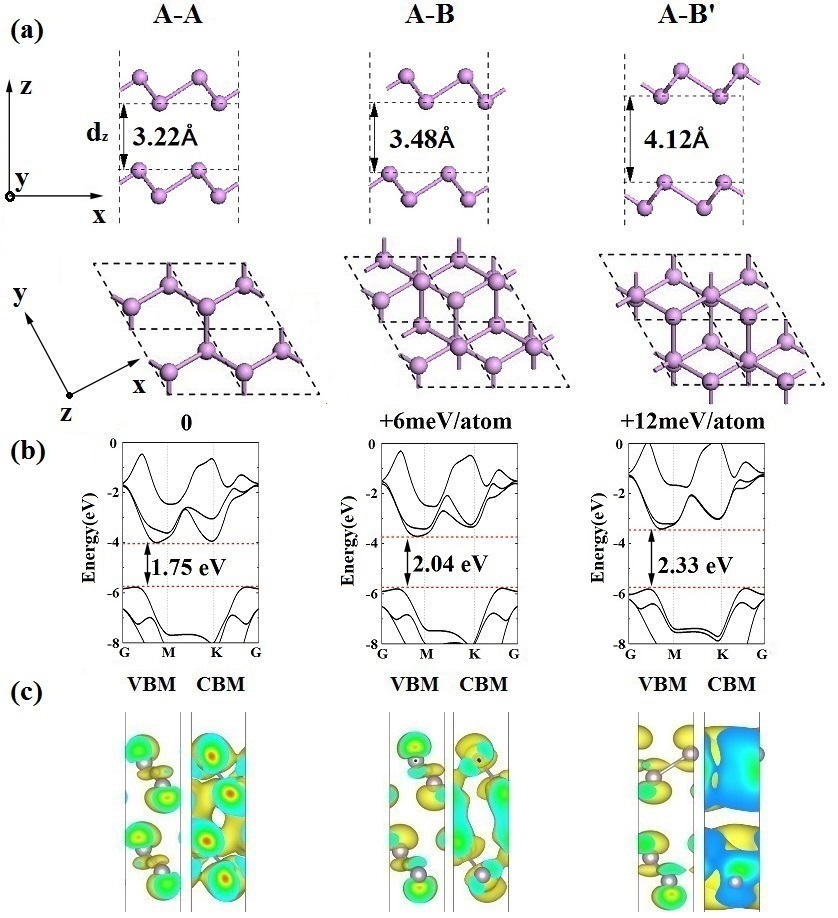}
\caption{(color online) Atomic geometry structures (a), band structures (b) and oribital probability density distribution of VBM and CBM (c) of A-A, A-B and A-B' stacking BLBP. Corresponding $E_{total}$, $E_{gap}$ and $d_z$ are also given. For $E_{total}$, the energy of A-A stacking is set to zero. For the oribital probability density distribution iso-surface plotting, the iso-value is 0.004 $e/au^{3}$.}
\end{figure}

The $\sigma$ bonding character and delocalized probability density distribution of CBM suggest that the vertical interlayer distance $d_z$ to be a key factor to determine the energy of CBM. In Fig.3 (b) we plot out $E_{total}$, CBM energy ($E_{CBM}$) and $d_z$ corresponding to lateral displacement in $x$ direction ($d_x$) of the top layer. One can see that the $E_{total}$, $E_{CBM}$ and $d_z$ follow the same trend correlated with $d_x$. $d_z$ changes remarkably from 3.2 {\AA} for A-A stacking to 4.1 {\AA} for A-B' stacking as shown in Fig.2 (a). The strongly configuration dependent interlayer distance can be understood by repulsive steric effects of the corrugated atomic layers as discussed by Liu et.al. \cite{25}. The lowest energy A-A stacking has the smallest dz because the protrudent atoms in the top layer sit in the vacancies of the bottom layer resulting in reduced repulsion. For A-B' stacking the protrudent atoms in top and bottom layers are face-to-face, leading to the largest repulsion and $d_z$.

\begin{figure}
\includegraphics[width=0.9\textwidth]{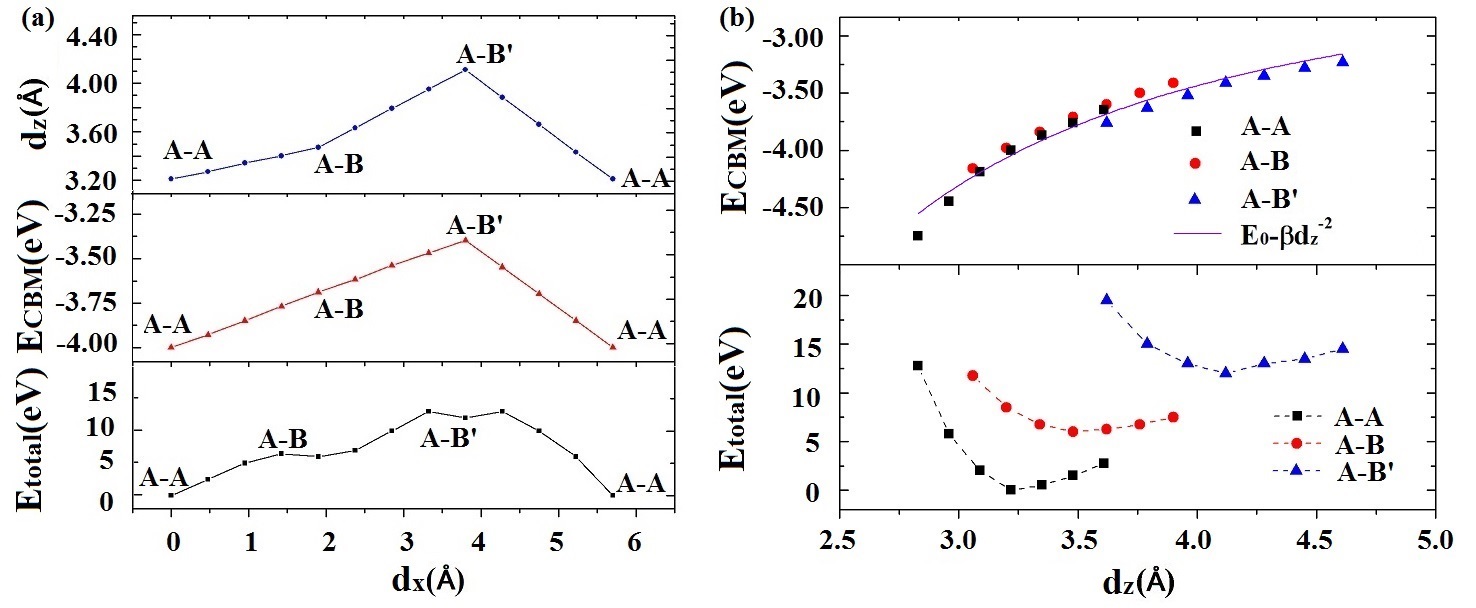}
\caption{(color online) (a) The change of $E_{total}$, $E_{CBM}$ and $d_z$ correlated with the lateral interlayer displacement $d_x$ on the route of A-A 每 A-B 每 A-B'每 A-A. (b) The change of $E_{total}$ and $E_{CBM}$ correlated with $d_z$ for the vertical interlayer displacement of A-A, A-B and A-B' stacking configurations. The fitting of $E_{CBM} = E_{0} - \beta d_{z}^{-2}$ is also shown.}
\end{figure}

The distinct variation of $E_{CBM}$ with changing of $d_z$ implies the vertical displacement may affect the electronic structure. In Fig.3 (b) we plot out $E_{CBM}$ and $E_{total}$ when we move the top layer BP vertically for different stacking configurations. We found that despite the different stacking configurations, $E_{CBM}$ follows quadratic curve equation which could be fit as: $E_{CBM} = E_{0} - \beta d_{z}^{-2}$, where $E_0$ and $\beta$ are two parameters (The values are given in Table S1). $E_0$ converges to $E_{CBM}$ for single layer BP when $d_z$ is infinite while $\beta$ is determined by the interlayer orbital hopping integral of CBM. The independence of $\beta$ on stacking configuration implys the interlayer interaction for different stacking is similar. This $d_{z}^{-2}$ variation is supported by tight-binding model presented in ref \cite{47}. Because the energy of VBM keeps as a constant, it is clear that $E_{gap}$ also follows the same $d_{z}^{-2}$ law with $E_{CBM}$. The variation of $E_{CBM}$ and $E_{gap}$ is as large as 1.5 eV within the vertical displacement of 1.8 {\AA} and total energy change of 20 meV.

\begin{figure}
\includegraphics[width=0.9\textwidth]{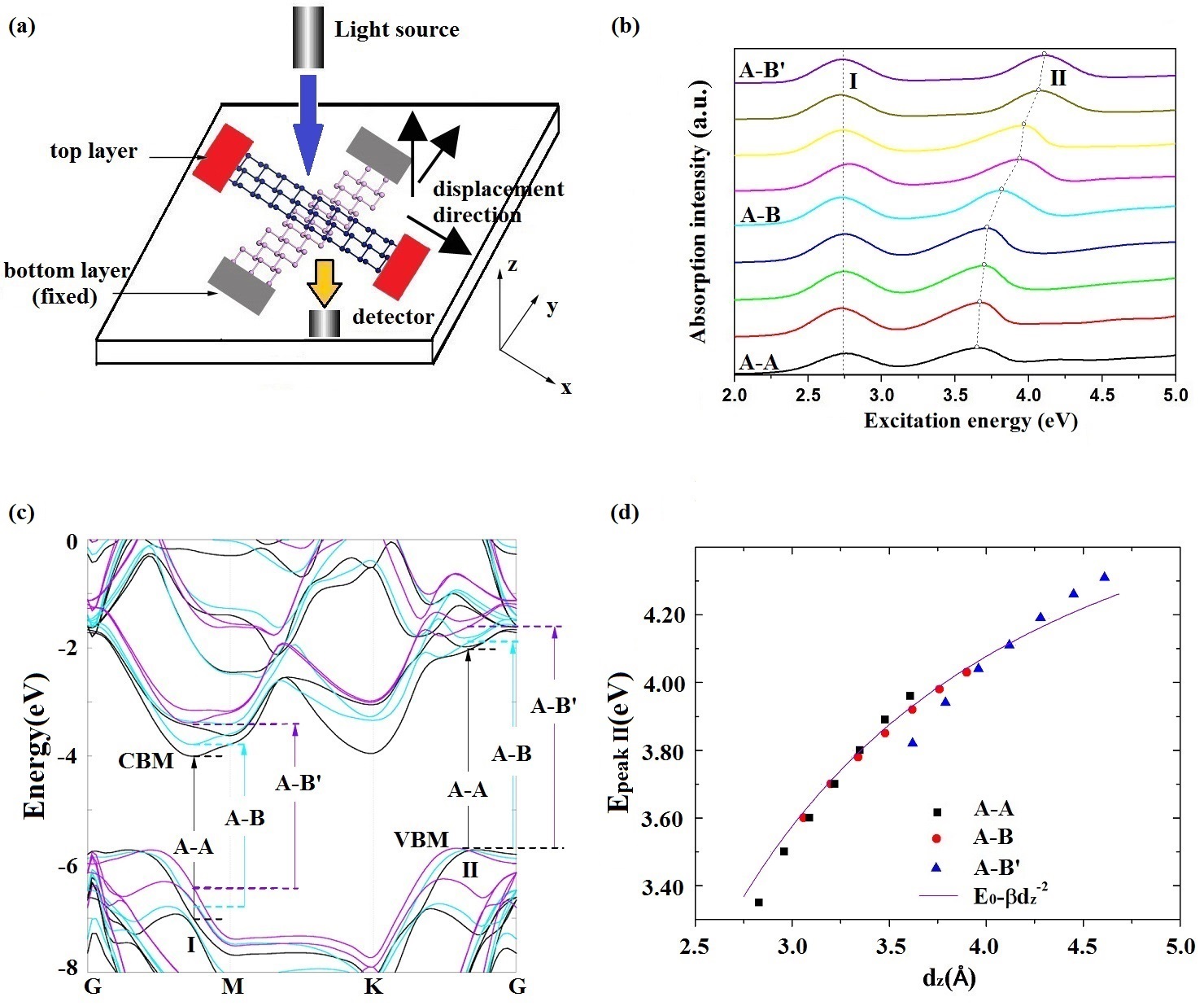}
\caption{(color online). (a) The schematic design diagram of the nano-scale displacement sensor based on BLBP. (b-c) The Absorption intensity (b) and band structure (c) of BLBP when the top layer is moved laterally from A-A (black line) through A-B (cyan line) to A-B' (purple line). (d) The change of $E_{Peak II}$ correlated with $d_z$ for the vertical interlayer displacement of A-A, A-B and A-B' stacking configurations. The fitting of $E_{Peak II} = E_{0} - \beta d_{z}^{-2}$ is also shown.}
\end{figure}

Because the electronic structure is very sensitive to the interlayer displacement, BLBP is proposed to be used as a nano-scale displacement sensor for weak-force systems. We provide a schematic design as shown in Fig.4 (a). During the measurement, the bottom layer is fixed while the top layer can be shifted laterally or vertically by a very weak force. The displacement is determined by measuring the electronic structure change. We propose to measure the electronic structure using photo absorption spectroscopy. Fig.4 (b) and (c) shows the calculated absorption intensity and the band structure change when the bottom layer is fixed and top layer is shifted laterally from A-A stacking to A-B' stacking. Two absorption peaks are obtained when the excitation energy is below 5 eV. Peak I at 2.75 eV which corresponds to the vertical transition from VB to CBM does not change with the top layer lateral displacement. This is because of this transition, the initial state in VB changes in the same trend with CBM as shown in Fig.4 (c). Peak II, which corresponds to the vertical excitation from VBM to CB, changes from 3.65 eV to 4.11 eV, showing significant sensitivity of top layer displacement because VBM remains constant and the final state in CB varies distinctly. (See the details of transitions I and II in supplementary materials.) Through the vertical displacement of different stacking configurations, we found the energy of peak II ($E_{Peak II}$) also follows the simple $d_{z}^{-2}$ trend as shown in Fig.4 (d). We propose that peak I can be used as a benchmark while the displacement could be deduced from the position of peak II. Especially, regardless of different stacking configurations, the vertical displacement could be easily obtained through the simple $E_{Peak II} \propto d_{z}^{-2}$ relationship. In addition, the BLBP electronic structure is also sensitive to the interlayer twist angle. (see the supplementary materials) Therefore, this schematic sensor can be upgraded to a four dimensional (three dimensions plus twist angle) displacement sensor in nano-scale.

\begin{figure}
\includegraphics[width=0.5\textwidth]{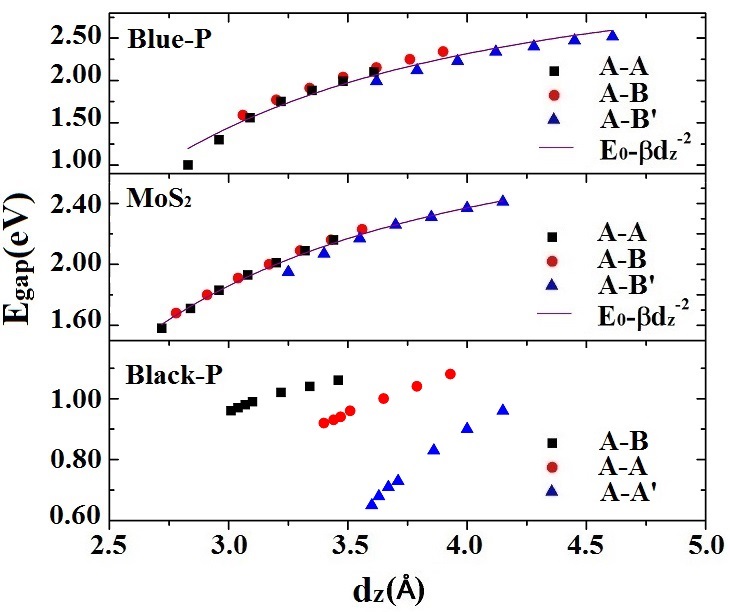}
\caption{(color online).The change of $E_{gap}$ with $d_z$ for the vertical interlayer displacement of A-A, A-B and A-B' stacking configurations for BLBP, bi-layer MoS$_2$ and black phosphrus. For BLBP and MoS$_2$, the fitting of $E_{gap} = E_{0} - \beta d_{z}^{-2}$ is also shown.}
\end{figure}

We propose this principle displacement sensor could be realized using different vertically stacked 2D bi-layer materials if they satisfy the following two requirements: (i) The atomic layer is corrugated so that the interlayer distance $d_z$ changes significantly with lateral displacement and twisting. (ii) The CB or VB has significant delocalized interlayer probability density distribution. Therefore the band structure of CB/VB changes remarkably with different $d_z$. We notice that analogous semiconductors as bi-layer MoS$_2$ and black phosphorus are found to be sensitive to the stacking configurations, which support our deduction \cite{25,48,49}. To convince this, we have calculated the $E_{gap}$ change versus $d_z$ of bi-layer MoS$_2$ and black phosphorus when the top layer is vertically displaced as shown in Fig.5. Appreciable $E_{gap}$ change can be observed for these three bi-layer materials. We found that $E_{gap}$ of bi-layer MoS$_2$ and BLBP, which are graphene-like 2D materials, follow the unified $d_{z}^{-2}$ law despite different stacking configuration. For black phosphorus, since atomic structure is different and the corrugation is very large (Fig. S4), the unified $d_{z}^{-2}$ law for different stacking configurations fails. Thus we conclude that both BLBP and bi-layer MoS$_2$ can be easily used to measure the $d_z$ change since $E_{Peak II} \propto d_{z}^{-2}$, however, BLBP has larger $\beta$ value (shown in Table S1) meaning the $E_{gap}$ tunability of BLBP is larger. Therefore we propose BLBP is a prominent candidate for displacement sensor for weak force system with high resolution.

In summary, using first principles calculations, we found the electronic structure of BLBP coupled through VdW interaction is sensitive to the interlayer displacement. Despite of different stacking configurations, the variation of band gap and interlayer distance $d_z$ follow a simple relationship as: $E_{Peak II} \propto d_{z}^{-2}$. This simple $d_{z}^{-2}$ law also works for other graphene-like corrugated bi-layer materials, such as MoS$_2$. Based on the tenability of electronic structure, we propose BLBP, as a representative of weakly coupled 2D semiconductor family, can be used as nano-scale displacement sensor.

This work is supported by National Key Basic Research Program (2011CB921404), by NSFC (21421063, 11322434), by CAS (XDB01020300), and by USTCSCC, SC-CAS, Tianjin and Shanghai Supercomputer Centers.



\end{document}


\title{Supplementary Materials for Nano-scale displacement sensing based on Van der Waals interaction}

\author{Lin Hu}%
\affiliation{ICQD/Hefei National Laboratory for Physical Sciences at Microscale, and Key Laboratory of Strongly-Coupled Quantum Matter Physics, Chinese Academy of Sciences, and Department of Physics, University of Science and Technology of China, Hefei, Anhui 230026, China}

\affiliation{Synergetic Innovation Center of Quantum Information $\&$ Quantum Physics, University of Science and Technology of China, Hefei, Anhui 230026, China}

\author{Jin Zhao}%
\email{zhaojin@ustc.edu.cn}
\affiliation{ICQD/Hefei National Laboratory for Physical Sciences at Microscale, and Key Laboratory of Strongly-Coupled Quantum Matter Physics, Chinese Academy of Sciences, and Department of Physics, University of Science and Technology of China, Hefei, Anhui 230026, China}

\affiliation{Synergetic Innovation Center of Quantum Information $\&$ Quantum Physics, University of Science and Technology of China, Hefei, Anhui 230026, China}

\author{Jinlong Yang}
\email{jlyang@ustc.edu.cn}
\affiliation{ICQD/Hefei National Laboratory for Physical Sciences at Microscale, and Key Laboratory of Strongly-Coupled Quantum Matter Physics, Chinese Academy of Sciences, and Department of Physics, University of Science and Technology of China, Hefei, Anhui 230026, China}

\affiliation{Synergetic Innovation Center of Quantum Information $\&$ Quantum Physics, University of Science and Technology of China, Hefei, Anhui 230026, China}

\date{\today}


\maketitle

\section{I. Vertical transitions corresponding to Peak I and Peak II.}

Peak I in Fig. 4(b) corresponds to a vertical transition from an initial state in VB to CBM. In Fig. S1 we plot out the orbital probability density distribution of the initial state and CBM. One can see that for three different stacking configurations, both the initial state and CBM show the bonding character of delocalized states in the interlayer region. Therefore their energy variations are almost the same. That is the reason that Peak I almost keeps constant corresponds to the displacement. Peak II in Fig.4 (b) corresponds to a vertical transition from VBM to a final state in CB. From Fig. S1 one can see that the final state contains the interlayer delocalized probability density and $\sigma$ bonding character while VBM does not. Therefore $E_{VBM}$ remains constant and $E_{Peak II}$ changes remarkably with the energy of the final state.

\begin{figure}
\includegraphics[width=0.30\textwidth]{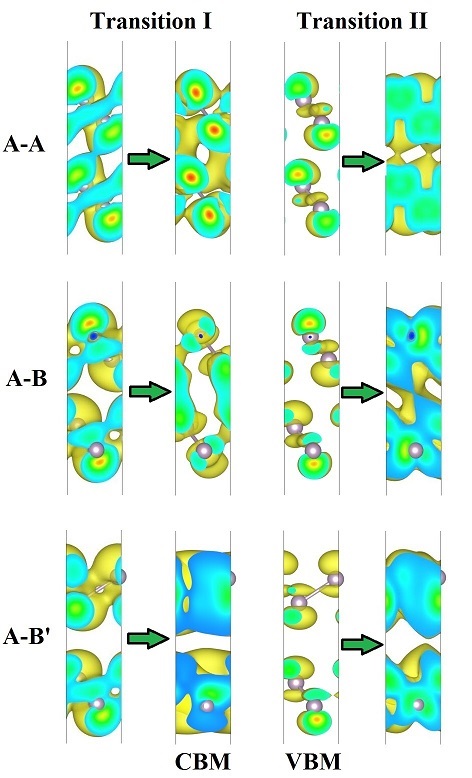}
\caption{(color online)The orbital probability density distributions of the initial and final states of vertical transitions corresponding to Peak I and Peak II for A-A, A-B and A-B' stacking BLBP.}
\end{figure}

\section{II. Fitting $E_{CBM}$ and $E_{Peak II}$ with quadratic equation}

Both $E_{CBM}$ and $E_{Peak II}$ follow the quadratic equation express as: $E = E_{0} - \beta d_{z}^{-2}$. The values of $E_{0}$ and $\beta$ are shown in Table S1. Here $E_{0}$ converges to the $E_{CBM}$/$E_{Peak II}$ value of single layer BP. $\beta$ is determined by the interlayer hopping integral of interacting orbitals, which is proportional to the wave function overlap. The $\beta$ value of $E_{CBM}$ is larger than $E_{Peak II}$, suggesting the hopping integral of CBM is larger than the final state of peak II in CB. For both $E_{CBM}$ and $E_{Peak II}$, $\beta$ is independent for different stacking configurations, which suggests that the interlayer hopping integral does not change much for different stacking configurations. This might be due to the delocalized distribution of the interacting orbitals. The $d_{z}^{-2}$ is described in tight-binding theory in ref\cite{1}.

\begin{table*}[bht]
  \tabcolsep 0.2 cm
  \caption{The values of $E_0$ and $\beta$ for $E_{CBM}$, $E_{Peak II}$ and $E_{gap}$.}
  \begin{tabular}{ccccc}
    \hline
                &                                   & $E_{CBM}$  & $E_{Peak II}$ & $E_{gap}$  \\
    \hline
    BLBP        & $E_{0}$(eV)                       & -2.31      & 4.80          & 3.4        \\
                & $\beta$(eV$\times$${\AA}^{2}$)    & 18.0       & 11.5          & 18.0       \\
    \hline
    Bi-layer    & $E_{0}$(eV)                       &            &               & 3.0        \\
    MoS$_2$     & $\beta$(eV$\times$${\AA}^{2}$)    &            &               & 10.5       \\
    \hline
  \end{tabular}
\end{table*}

\section{III. The electronic structure dependence on interlayer twisting.}

The electronic structure of BLBP is also sensitive to the interlayer twisting. We have investigated the band gap change for the A-A stacking when the top layer is twisted around $z$ axis shown in Fig. S2. The band gap varies from 1.75 eV to 2.34 eV when the twist angle $\theta$ changes from $0^{\circ}$ to $180^{\circ}$. The total energy is within 13 meV. The interlayer distance $d_z$ changes from 3.22 {\AA} to 4.12 {\AA}. The lowest energy configuration corresponding to $\theta = 0^{\circ}$ has the smallest $d_z$, in which the CBM is stabilized to the lowest energy because of its $\sigma$ bonding character.

\begin{figure}
\includegraphics[width=0.66\textwidth]{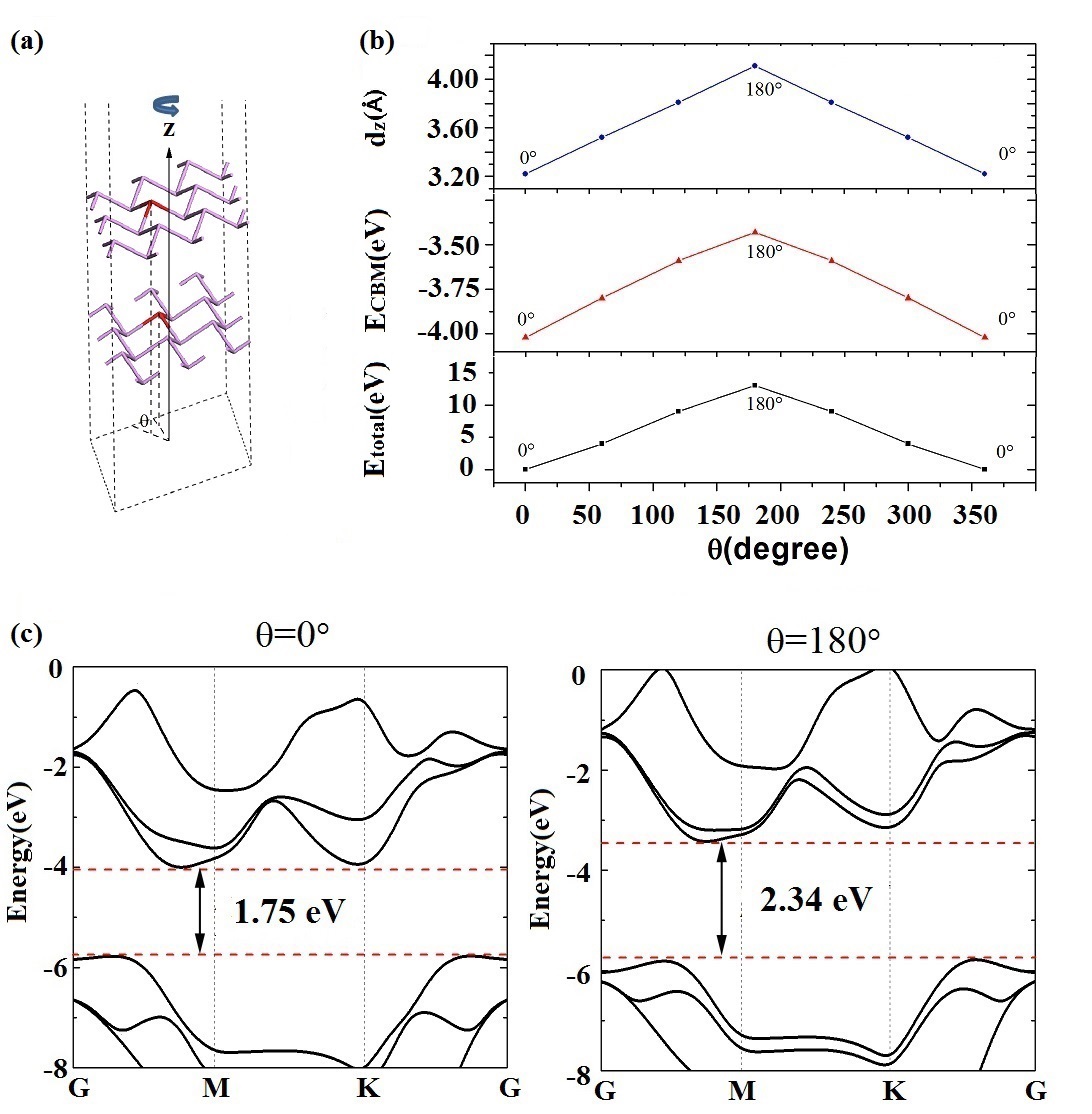}
\caption{(color online)The electronic structure dependence on interlayer twisting for A-A stacking BLBP. (a) The schematic diagram of interlayer twisting. (b) The change of $E_{total}$, $E_{CBM}$ and $d_z$ with twisting angle $\theta$. (c) The plotting of band structure corresponding to $\theta = 0^{\circ}$ and $\theta = 180^{\circ}$.}
\end{figure}

The absorption intensity of different twisting angle is also investigated. The absorption intensity for A-A stacking BLBP with different interlayer twist angle is shown in Fig. S3(a). The lowest two peaks corresponding to two direct excitation involves CBM and VBM as shown in Fig. S3(b). Peak I corresponding to the vertical excitation to CBM does not change with the twist angle. Peak II corresponding to the vertical excitation from VBM changes from 3.65 eV to 4.18 eV when $\theta$ changes from $0^{\circ}$ to $180^{\circ}$.

\begin{figure}
\includegraphics[width=0.75\textwidth]{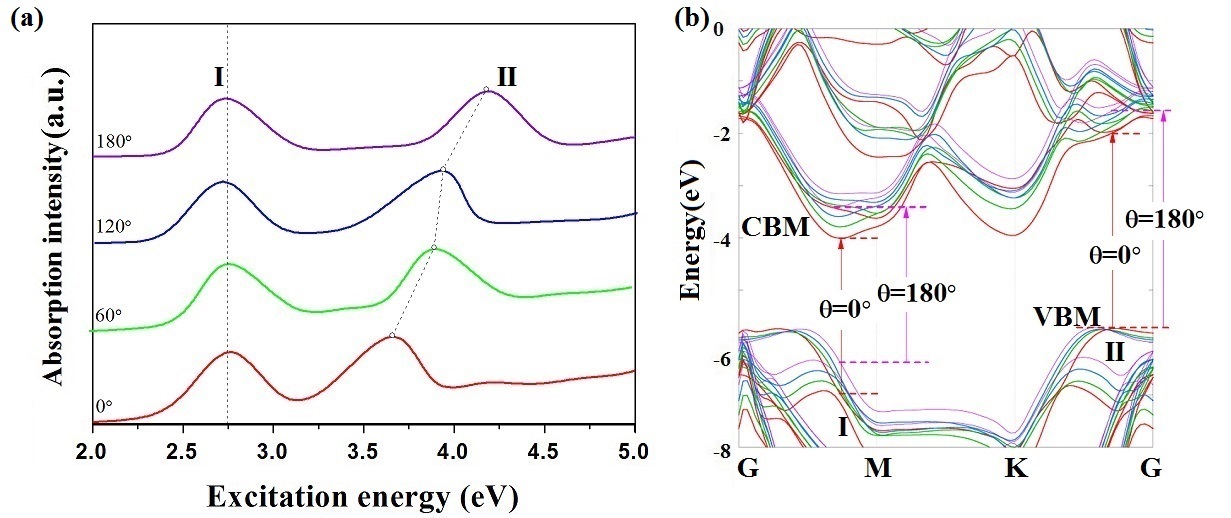}
\caption{(color online)The Absorption intensity (a) and band structure (b) of A-A stacking BLBP when the interlayer twisting angle $\theta$ changes from $0^{\circ}$ (red line) to $180^{\circ}$ (purple line). The two absorption peaks below 5 eV are marked as peaks I and II in (a). The corresponding vertical transitions for $\theta = 0^{\circ}$ and $\theta = 180^{\circ}$ are indicted by red and purple arrows in (b).}
\end{figure}

\section{IV. Comparation with bi-layer MoS$_2$ and black phosphorus}

We have compare BLBP with other corrugated bi-layer materials MoS$_2$ and black phosphorus. Fig. S4 shows the top view and side view of different stacking configurations of MoS$_2$ and black phosphorus. One can see that MoS$_2$ has similar structure with blue phosphorus (Fig. S4(a)), which belongs to the graphene-like structures. As shown in Fig. 5, similar with BLBP, $E_{gap}$ of bi-layer MoS$_2$ can be fitted with $E = E_{0} - \beta d_{z}^{-2}$. In contrast, the corrugation of black phosphorus is very large as shown in Fig. S4(b), therefore the unified $E_{gap} \propto d_{z}^{-2}$ trend for different stacking configuration fails.

\begin{figure}
\includegraphics[width=0.86\textwidth]{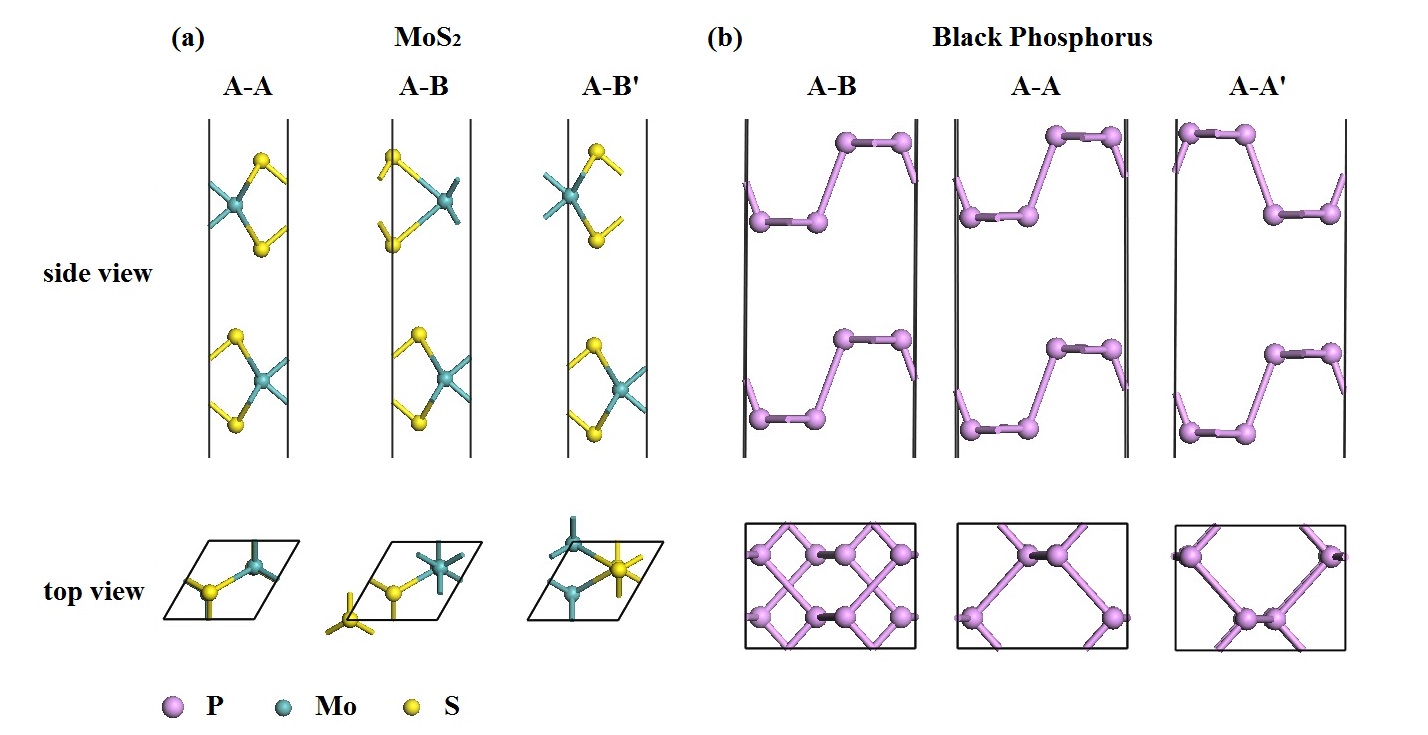}
\caption{(color online)The side view and top view of atomic structure of A-A, A-B, and A-B' stacking MoS$_2$ and black phosphorus.}
\end{figure}